\documentclass[aps,prl,twocolumn,showpacs,superscriptaddress,groupedaddress]{revtex4}  
\usepackage{graphicx}  
\usepackage{dcolumn}   
\usepackage{bm}        
\usepackage{amssymb}   
\usepackage{amsmath}
\newcommand{\vect}[1]{\boldsymbol{#1}}
\usepackage{lipsum}

\begin{document}

\widetext
\leftline{Version xx as of \today}

\title{Internal noise driven generalized Langevin equation from a nonlocal continuum model}

\author{Saikat~Sarkar} \affiliation{Computational Mechanics Lab., Indian Institute of Science, Bangalore 560012, India}
\author{Shubhankar~Roy~Chowdhury} \affiliation{Computational Mechanics Lab., Indian Institute of Science, Bangalore 560012, India}
\author{Debasish~Roy} \affiliation{Computational Mechanics Lab., Indian Institute of Science, Bangalore 560012, India}
\author{Ram Mohan~Vasu} \affiliation{Instrumentation and Applied Physics, Indian Institute of Science, Bangalore 560012, India}

\noaffiliation
\vskip 0.25cm

\date{\today}

\begin{abstract}
Starting with a micropolar formulation, known to account for nonlocal microstructural effects at the continuum level,
a generalized Langevin equation (GLE) for a particle, describing the predominant motion of a localized region through a single displacement
degree-of-freedom (DOF), is derived. The GLE features a memory dependent multiplicative or internal noise, which appears upon recognising that
the micro-rotation variables possess randomness owing to an uncertainty principle. Unlike its classical version, the new GLE qualitatively
reproduces the experimentally measured fluctuations
in the  steady-state mean square displacement of scattering centers in a polyvinyl alcohol slab. The origin of the fluctuations is traced to nonlocal spatial interactions within the continuum.
A constraint equation, similar to a fluctuation dissipation theorem (FDT), is shown to statistically relate the internal noise to the other parameters in the GLE.
\end{abstract}

\maketitle

As the space/time length scale in the forcing or the triggered deformation mechanism becomes comparable
with the internal length scale of the material, the system response at the macro-continuum scale is significantly influenced by the material
microstructure. This renders the classical continuum hypothesis of strictly local interactions, or its possible extension using adiabatic continuation arguments, untenable
in the modeling of such response. Non-local modeling techniques such as micropolar \cite{eringen1},
micromorphic \cite{eringen1} or gradient \cite{grad1} theories, which aim at incorporating long-range inter-particle interactions, bring forth
microstructural effects by introducing material length scales in the constitutive formulation.
In materials like polymers, granular solids etc. where the length scale is of the macroscopic order,
long-range effects could predominate and, in such cases, predictions through nonlocal models, unlike the the classical continuum model, are in closer conformity with
experimental observations.
If a continuum model can be replaced by a collection of harmonic oscillators and the focus is on the motion of a small region represented by
a particle (an oscillator) with a single predominant translational degree-of-freedom (DOF), one arrives at a generalized Langevin equation (GLE)
for the DOF after including the coupling effects from the neighboring oscillators \cite{zwanzig}. The GLE, an expedient modeling tool
that replaces the infinite dimensional continuum, is widely used in areas such as soft condensed matter physics and cell biology. Despite the correspondence
as above between the GLE and the continuum, standard forms of the GLE do not include length scale information characteristic of nonlocal
continuum theories and are therefore not equipped to describe the physically relevant microstructural effects.

\par
\begin{figure}[!htp]
    \begin{center}
            \label{1}
            \includegraphics[width=0.4\textwidth]{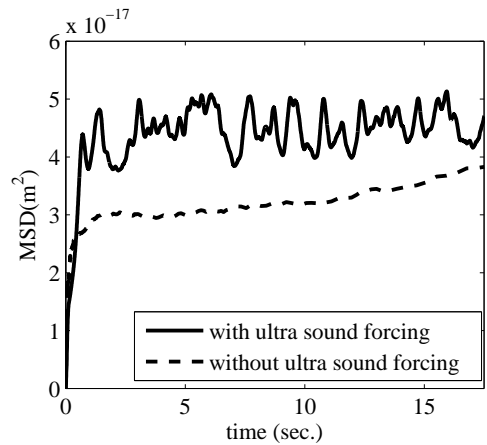}
    \end{center}
    \caption{%
Experimental plots of MSDs
     }%
     \label{exp_GLE}
\end{figure}

\begin{figure}[!htp]
    \begin{center}
            \label{1}
            \includegraphics[width=0.4\textwidth]{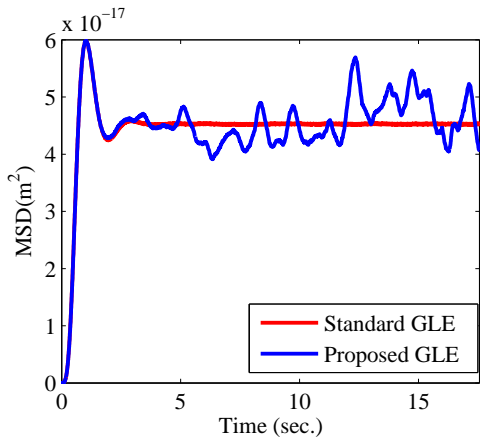}
    \end{center}
    \caption{%
Simulation via standard and proposed GLE
     }%
     \label{simulation_GLE}
\end{figure}

 This work is partly motivated by the experimental data shown in Fig.\ref{exp_GLE} (adapted from Fig. 6 of \cite{r1}).
 These correspond to the mean square displacement (MSD) plots of temperature-induced Brownian particles in a polyvinyl alcohol (PVA) slab, extracted through light scattering studies: one corresponding to a local region marked by a focussed high-frequency ultrasound force and the other a global measurement valid also for the region. In the steady-state, the MSD in the ultrasound forcing case shows
 significant fluctuations, though over a much lower frequency band. These fluctuations are absent when the ultrasound forcing is switched off.
 Attempts at numerical simulations using the standard GLE under the known ultrasound forcing fail to reproduce such fluctuations (see Fig.\ref{simulation_GLE}),
the origin of which is believed to be in the long-range microstructural effects.

\par In this letter, we aim to arrive at a GLE that carries such microstructural information. A key aspect is an internal noise term in the GLE, which
reflects an inherent randomness in the micro-rotations consequent to an uncertainty relation
involving two strain operators of which only one contains micro-rotational information. Through the internal noise, evolution of the microstructural
interactions, as manifested in the continuum scale, is characterized. Numerical simulations of the new GLE is shown to capture the steady-state fluctuations in the MSD.
Extending Kubo's second fluctuation dissipation theorem (FDT)\cite{kubo}, a constraint equation, relating the damping memory kernel and intensities of both the initial thermal fluctuation and the internal noise, is derived.

\par\emph{Discrete Hamiltonian from a micropolar perspective.---} This work is founded on the premise that the adoption of micropolarity in the medium suffices
to bring in the microstructural information.
Accordingly, in pursuance of the micropolar continuum theory, the deformation kinematics requires each material point to have
micro-rotation DOFs in addition to the standard translational DOFs.
The total mechanical energy functional ($\Pi$) for a geometrically nonlinear micropolar body of material volume $V_0$ may be represented as,
$\Pi=\frac{1}{2}\int_{{{V}_{0}}}{{{\rho }_{0}}\left( \vect{v}.\vect{v}+J\vect{\omega} .\vect{\omega} \right)d\vect{V}}+\int_{{{V}_{0}}}{{{\rho }_{0}}\psi d\vect{V}}$.
Here $\vect{v}$ is the linear velocity, $\vect{\omega}$ the spin velocity, $\rho_0$ the mass density, $\rho_0 J \vect{I}$ the micro-inertia tensor
and $\psi$ the specific free energy potential, all represented in Lagrangian coordinates.
Assuming micro-rotations to be small, possibly an order smaller than the translation DOFs, a consistent spatial discretization
of $\Pi$, as elaborated in the supplementary material\cite{supplementary} with the details of the symbols used, leads to the Hamiltonian ($H$) corresponding to a discrete representation of the body.
In the discrete body, surrounded by a set of bath particles is a system particle whose predominant translation DOF, also called the system DOF, is described by the GLE that we intend to arrive at.
 Indeed, except for the system DOF, all other DOFs appearing in $H$ are henceforth referred to
as the bath DOFs. Accordingly, we split the Hamiltonian as $H=H_s+H_b$,
where ${{H}_{s}}=\frac{{{p}_{11}^{2}}}{2{{m}_{11}}}+{{{w}}_{1111}}{{u}_{11}^{2}}$ does not explicitly involve the bath DOFs and
\[\begin{split} &{{H}_{b}}=\frac{p_{21 }^{2}}{2{{m}_{21}}}+\frac{p_{31 }^{2}}{2{{m}_{31}}}+\sum\limits_{\alpha =2}^{N}{\frac{p_{j\alpha }^{2}}{2{{m}_{j\alpha }}}}+
\sum\limits_{\alpha =2}^{N}{\frac{p_{\theta j\alpha }^{2}}{2{{I}_{j\alpha }}}}+\\&\sum\limits_{\alpha ,\beta =2}^{N}{{{{{w}}}_{ij\alpha \beta }}{{u}_{j\alpha }}{{u}_{i\beta }}}+
	\sum\limits_{\beta =2}^{N}{{{{{w}}}_{i11\beta }}{{u}_{11}}{{u}_{i\beta }}}+\\	
	&{\sum\limits_{\alpha,\gamma =1}^{N}{{\tilde{w}_{i\alpha \gamma }}{u_{j\alpha }}\left(\left(\theta_{j\gamma}\theta_{i\gamma}+2\text{S}\left(\theta_{k\gamma}\right)_{k\ne j,i}\right)_{i\ne j}\right)}}+
	\\&{\sum\limits_{\alpha,\gamma =1}^{N}{{\tilde{w}_{i\alpha \gamma }}{u_{j\alpha }}\left(1-\sum\limits_{k\ne i}{\theta_{k\gamma}^2}\right)_{i=j}}}\\
   \end{split}\]
Here $i, j\in{\left\{1,2,3\right\}}$ are the indices denoting the three cartesian coordinates. Summations over the repeated indices ($i$, $j$)  are implied.
$m_{k\alpha}$ and $I_{k\alpha}$ are respectively the mass and mass-moment of inertia of the $\alpha^{\text{th}}$ particle. The coordinate index  $k$, fictitiously
introduced in $m$ and $I$, serves to maintain indicial consistency with the vectors $\vect{u}$, $\vect{\theta}$ and $\vect{p}$ appearing in the expressions and helps
in using Einstein's summation convention in $k$.
$p_{i\alpha}$, $p_{\theta i\alpha}$ are respectively the linear and angular momentum components and
$u_{i\alpha}$, $\theta_{i\alpha}$ the displacement and micro-rotation components, all evaluated at the $\alpha^{\text{th}}$ particle in the $i^{\text{th}}$ direction. Of
specific interest are $u_{11}$ and $p_{11}$, the displacement and the momentum components of the system particle in the required direction, for which the GLE will be written.
For $\left(j,i\right)=\left(1,2\right), \left(2,3\right), \left(3,1\right)$, $\text{S}=1$ and for other combinations of $\left(j,i\right)$, $\text{S}=-1$. $N$ is the number of particles in the discrete body.
\par \emph{Formulation of the new GLE.---}
Using the discrete Hamiltonian $H$, the governing dynamics for the system and bath variables, described in terms of the displacement DOFs and the momenta, are obtained through Hamilton's equations; details of the derivation are provided in \cite{supplementary}.
The equations of motion, in a compact form, are given by Eqn. (\ref{5}) and (\ref{6}).
\begin{equation}\label{5}\begin{split}&\frac{d}{dt}\left\{ \begin{matrix}
   {{u}_{11}}  \\
   {{p}_{11}}  \\
\end{matrix} \right\}=\left[ \begin{matrix}
   0 & {1}/{{{m}_{11}}}\;  \\
   {{{\hat{w}}}_{1111}} & 0  \\
\end{matrix} \right]\left\{ \begin{matrix}
   {{u}_{11}}  \\
   {{p}_{11}}  \\
\end{matrix} \right\}+\boldsymbol{\Upsilon}\left\{ \begin{matrix}
   {{{\hat{u}}}_{\theta }}  \\
   {{{\hat{p}}}_{u\theta }}  \\
\end{matrix} \right\}\end{split}\end{equation} 	
where

$\boldsymbol{\Upsilon}=\left[ \begin{matrix}
   {{\mathbf{0}}_{1\times \left( 6N-1 \right)}} & {{\mathbf{0}}_{1\times \left( 6N-1 \right)}}  \\
   {{\mathbf{w}}_{1\times \left( 6N-1 \right)}} & {{\mathbf{0}}_{1\times \left( 6N-1 \right)}}  \\
\end{matrix} \right]$ and
\begin{equation}\label{6}\begin{split}\frac{d}{dt}\left\{ \begin{matrix}
   {{{\hat{u}}}_{\theta }}  \\
   {{{\hat{p}}}_{u\theta }}  \\
\end{matrix} \right\}=&\left[ \begin{matrix}
   {{\mathbf{0}}_{\left( 6N-1 \right)\times \left( 6N-1 \right)}} & \mathbf{\Lambda }  \\
   \mathbf{K} & {{\mathbf{0}}_{\left( 6N-1 \right)\times \left( 6N-1 \right)}}  \\
\end{matrix} \right]\left\{ \begin{matrix}
   {{{\hat{u}}}_{\theta }}  \\
   {{{\hat{p}}}_{u\theta }}  \\
\end{matrix} \right\}+\\&{{u}_{11}}\left\{ \begin{matrix}
   {{\mathbf{0}}_{\left( 6N-1 \right)\times 1}}  \\
   \vect{g}  \\
\end{matrix} \right\}\end{split}\end{equation}
${{\hat{u}}_{\theta }}={{\left\{ \begin{matrix}
   {\hat{u}}^{\text{T}} & {\theta}^{\text{T}}   \\
\end{matrix} \right\}}^{\text{T}}}={{\left\{ \begin{matrix}
   {{u}_{21}} & ... & {{u}_{3N}} & {{\theta }_{11}} & ... & {{\theta }_{3N}}  \\
\end{matrix} \right\}}^{\text{T}}}$consists of displacements and micro-rotations of the bath variables and  ${{\hat{p}}_{u\theta }}={{\left\{ \begin{matrix}
   {{p}_{21}} & ... & {{p}_{3N}} & {{p}_{\theta 11}} & ... & {{p}_{\theta 3N}}  \\
\end{matrix} \right\}}^{\text{T}}}$, the corresponding linear and angular momenta. $\boldsymbol{\Lambda}$ is a diagonal matrix with nonzero entries $\left\{ \begin{matrix}
   {1}/{{{m}_{21}}}\; & ... & {1}/{{{m}_{3N}}}\; & {1}/{\left({I}_{11}\right)}\; & ... & {1}/{\left({I}_{3N}\right)}\;  \\
\end{matrix} \right\}$.
$\mathbf{K}=\left[ \begin{matrix}
   {{\mathbf{K}}_{1}} & {{\mathbf{K}}_{2}}+{{u}_{11}}{{\mathbf{K}}_{3}}  \\
\end{matrix} \right]$, ${{\mathbf{K}}_{2}}=\left[ \begin{matrix}
   \mathbf{A}  \\
   {{\mathbf{0}}_{3N\times 3N}}  \\
\end{matrix} \right]$and ${{\mathbf{K}}_{3}}=\left[ \begin{matrix}
   {{\mathbf{0}}_{\left( 3N-1 \right)\times 3N}}  \\
   \mathbf{B}  \\
\end{matrix} \right]$.  ${{\mathbf{K}}_{1}}$, $\mathbf{A}$ and $\mathbf{B}$ are constant matrices of dimensions
$\left( 6N-1 \right)\times \left( 3N-1 \right)$, $\left( 3N-1 \right)\times 3N$ and $3N\times 3N$respectively. $\vect{g}$ is a $\left( 6N-1 \right)$ dimensional constant vector.
$\mathbf{0}_{m\times n}$ designates a zero matrix of dimension $m\times n$ and $\mathbf{w}_{1\times \left( 6N-1 \right)}$ in $\boldsymbol{\Upsilon}$ is a constant matrix.
Denoting $\left\{\begin{matrix}{\hat{u}_\theta}^{\text{T}} & {\hat{p}_{u\theta}}^{\text{T}}\end{matrix}\right\}^\text{T}$ as $Y_b$, Eqn. (\ref{6}) may be written as:
\begin{equation}\label{7}\frac{d{{Y}_{b}}}{dt}=\mathbf{\bar{K}}{{Y}_{b}}+{{u}_{11}}{{\left\{ \begin{matrix}
   {{\mathbf{0}}_{\left( 6N-1 \right)\times 1}} & \vect{g}+\mathbf{B}\theta   \\
\end{matrix} \right\}}^{\text{T}}}\end{equation}
where $$\mathbf{\bar{K}}=\left[ \begin{matrix}
   {{\mathbf{0}}_{\left( 6N-1 \right)\times \left( 6N-1 \right)}} & \mathbf{\Lambda }  \\
   \left[ \begin{matrix}
   {{\mathbf{K}}_{1}} & {{\mathbf{K}}_{2}}  \\
\end{matrix} \right] & {{\mathbf{0}}_{\left( 6N-1 \right)\times \left( 6N-1 \right)}}  \\
\end{matrix} \right].$$
Multiplying Eqn.(\ref{7})  with $\exp \left( -\mathbf{\bar{K}}t \right)$ and integrating over $\left[ 0,t \right]$, we get the following implicit expression:

\begin{equation}\label{8}\begin{split}&{{Y}_{b}}\left( t \right)=\exp \left( \mathbf{\bar{K}}t \right){{Y}_{b}}\left( 0 \right)+\exp \left( \mathbf{\bar{K}}t \right)
.\\&\int\limits_{0}^{t}{\exp \left( -\mathbf{\bar{K}}s \right){{u}_{11}}\left( s \right){{\left\{ \begin{matrix}
   {{\mathbf{0}}_{\left( 6N-1 \right)\times 1}} & \vect{g}+\mathbf{B}\theta \left( s \right)  \\
\end{matrix} \right\}}^{\text{T}}}ds}\end{split}\end{equation}

Integration by parts on the term $\int\limits_{0}^{t}{\exp \left( -\mathbf{\bar{K}}s \right){{u}_{11}}\left( s \right){{\left\{ \begin{matrix}
   {{\mathbf{0}}_{\left( 6N-1 \right)\times 1}} & \vect{g}  \\
\end{matrix} \right\}}^{\text{T}}}ds}$ of Eqn.(\ref{8}) would lead to an equivalent representation given in Eqn.(\ref{9}).
\begin{equation}\label{9}\begin{split}
  &{{Y}_{b}}\left( t \right)=\exp \left( \mathbf{\bar{K}}t \right){{Y}_{b}}\left( 0 \right)+{{{\mathbf{\bar{K}}}}^{-1}}{{u}_{11}}\left( t \right){{\left\{ \begin{matrix}
   {{\mathbf{0}}_{\left( 6N-1 \right)\times 1}}& \vect{g}\end{matrix} \right\}}^{\text{T}}}-\\
&\exp \left( \mathbf{\bar{K}}t \right){{{\mathbf{\bar{K}}}}^{-1}}{{u}_{11}}\left( 0 \right){{\left\{ \begin{matrix}{{\mathbf{0}}_{\left( 6N-1 \right)\times 1}}& \vect{g}\end{matrix} \right\}}^{\text{T}}}+\\
& \exp \left( \mathbf{\bar{K}}t \right)\int\limits_{0}^{t}{\exp \left( -\mathbf{\bar{K}}s \right){{{\mathbf{\bar{K}}}}^{-1}}{{{\dot{u}}}_{11}}\left( s \right)
{{\left\{ \begin{matrix}{{\mathbf{0}}_{\left( 6N-1 \right)\times 1}}& \vect{g}\end{matrix} \right\}}^{\text{T}}}ds}+  \\
&\exp \left( \mathbf{\bar{K}}t \right)\int\limits_{0}^{t}{\exp \left( -\mathbf{\bar{K}}s \right){{u}_{11}}\left( s \right){{\left\{ \begin{matrix}
{{\mathbf{0}}_{\left( 6N-1 \right)\times 1}} & \mathbf{B}\theta \left( s \right)  \\
\end{matrix} \right\}}^{\text{T}}}ds}
\end{split}\end{equation}
Substituting
${{Y}_{b}}\left( t \right)$, as in Eqn.(\ref{9}), into Eqn.(\ref{5}), we obtain the governing equations of motion (Eqn.(\ref{10})) for the system variables with the micro-rotation DOFs still tagged on.
\begin{equation}\label{10}
\begin{split}
 &\frac{d}{dt}\left\{ \begin{matrix}
   {{u}_{11}}  \\
   {{p}_{11}}  \\
\end{matrix} \right\}=\left[ \begin{matrix}
   0 & {1}/{{{m}_{11}}}\;  \\
   \left( {{{\hat{w}}}_{1111}}+{{{\overset{\lower0.5em\hbox{$\smash{\scriptscriptstyle\smile}$}}{w}}}_{0}}\right)&0\\\end{matrix}\right]\left\{\begin{matrix}{{u}_{11}}\\{{p}_{11}}\\\end{matrix}\right\}+\\
 &\boldsymbol{\Upsilon}
\int\limits_{0}^{t}{\exp\left(-\mathbf{\bar{K}}\left(s-t\right)\right){{{\mathbf{\bar{K}}}}^{-1}}{{{\dot{u}}}_{11}}\left(s\right)
\left\{\begin{matrix}{{\mathbf{0}}_{\left(6N-1\right)\times1}}\\\vect{g}\\\end{matrix}\right\}ds}+\\
 & \boldsymbol{\Upsilon}
   \left\{\exp\left(\mathbf{\bar{K}}t\right)\left({Y_b\left({0}\right)-\mathbf{\bar{K}}^{-1}u_{11}\left({0}\right)\left\{\begin{matrix}
   {\mathbf{0}_{\left(6N-1\right)\times1}}\\\vect{g}\end{matrix}\right\}}\right)\right\}+\\
   &\boldsymbol{\Upsilon}\int\limits_0^t{\exp\left(-\mathbf{\bar{K}}\left(s-t\right)\right)u_{11}\left(s\right)\left\{\begin{matrix}
   {\mathbf{0}_{\left(6N-1\right)\times1}}\\{\mathbf{B}\boldsymbol{\theta}\left(s\right)}\end{matrix}ds\right\}}
 \end{split}
 \end{equation}
The constant element ${{\overset{\lower0.5em\hbox{$\smash{\scriptscriptstyle\smile}$}}{w}}_{0}}$ in Eqn.(\ref{10}) additionally contributes to the stiffness due to micropolarity.
This constant is the second element of the two-dimensional vector
$\boldsymbol{\Upsilon}\mathbf{\bar{K}}^{-1}\left\{\begin{matrix}{\mathbf{0}_{\left(6N-1\right)\times1}}\;\;\vect{g}\end{matrix}\right\}^\text{T}$.

Our interest is in deriving a GLE for the system DOF $u_{11}$, which requires eliminating the micro-rotation DOFs from Eqn.(\ref{10}). Inherent
configurational uncertainty of the microstructure and its time evolution would seem to imply that the micro-rotation DOFs are treated as stochastic processes.
A further justification of this viewpoint is provided through the noncommutativity of the symmetrized polar and nonpolar operators,
$\hat{\vect{E}}_p=\frac{1}{2}\left(\tilde{\vect{E}}_{p}+\tilde{\vect{E}}_{p}^\text{T}\right)$ and
$\hat{\vect{E}}_{np}=\frac{1}{2}\left(\tilde{\vect{E}}_{np}+\tilde{\vect{E}}_{np}^\text{T}\right)$, wherein
${\tilde{\vect{E}}}_p=\vect{R}^\text{T}\vect{F}$ and ${\tilde{\vect{E}}}_{np}=\vect{F}^\text{T}\vect{F}$ are
respectively considered to be measures of polar and nonpolar strains. $\vect{R}$ and $\vect{F}$ are micro-rotation and deformation gradient tensors respectively \cite{supplementary}. Specifically, we can arrive at the following
Robertson-Schrodinger uncertainty \cite{robertson} relation involving these two operators,
\begin{equation}\label{uncer}
\begin{split}
&\sigma_{{\hat{\vect{E}}}_{p}}\sigma_{{\hat{\vect{E}}}_{np}}\ge \frac{1}{4}\left(\begin{split}&\left|\left\langle{\left[\hat{\vect{E}}_p\text{,}\hat{\vect{E}}_{np}\right]}\right\rangle\right|^2+\\
&\left|\left\langle{\left\{\hat{\vect{E}}_p-\left\langle{\hat{\vect{E}}_p}\right\rangle\text{,}\hat{\vect{E}}_{np}-\left\langle{\hat{\vect{E}}_{np}}\right\rangle\right\}}\right\rangle\right|^2\end{split}\right)\\
\end{split}
\end{equation}
${\left[\vect{A}\text{,}\vect{B}\right]}$ is the commutator and ${\left\{\vect{A}\text{,}\vect{B}\right\}}$
the anti-commutator of the operators $\vect{A}$ and $\vect{B}$. For a given symmetric matrix operator
$\vect{C}$, $\left\langle{\vect{C}}\right\rangle=\int_\Omega{\vect{f}(\vect{X};\vect{X'})^\text{T}\vect{C}(\vect{X'})\vect{f}(\vect{X};\vect{X'})}d\vect{X'}$ defines
the mean of $\vect{C}$ with respect to $\vect{f}$.  $\vect{f}(\vect{X};\cdot)$ is a compactly supported vector valued function of $\vect{X'}$ with an arbitrary support
$\Omega$ containing $\vect{X}$ and must be normalized, i.e. $||\vect{f}||^2=\int_\Omega{\vect{f}^\text{T}\vect{f}d\vect{X'}}=1$. Thus $\vect{f}^\text{T}\vect{f}$ is interpretable
as the density associated with a probability measure. Here, for instance, we choose
$\vect{f}=\frac{\vect{X}-\vect{X'}}{||\vect{X}-\vect{X'}||}$, a normalized line segment. Such a mean, though a scalar quantity, should be interpreted as an operator wherever appropriate.
 $\sigma_{\vect{A}}=\left\langle{\left({\vect{A}}-\left\langle{\vect{A}}\right\rangle\right)^\text{T}\left({\vect{A}}-\left\langle{{\vect{A}}}\right\rangle\right)}\right\rangle$
 is the variance (an uncertainty measure) associated with $\vect{A}$.
 In terms of $\vect{F}$ and $\vect{R}$, the inequality (\ref{uncer}) may be recast as,
 \begin{equation}\label{uncer2}
\begin{split}
\sigma_{{\hat{\vect{E}}}_p}\sigma_{{\hat{\vect{E}}}_{np}}\ge
\frac{1}{4}\left|\left\langle{\left(\vect{R}^\text{T}\vect{F}+\vect{F}^\text{T}\vect{R}\right)\left(\vect{F}^\text{T}\vect{F}
-\left\langle{\vect{F}^\text{T}\vect{F}}\right\rangle\right)}\right\rangle\right|^2\end{split}
\end{equation}
 The uncertainty relation is nontrivial only if the RHS of the inequality (\ref{uncer2}) is strictly greater than zero and this is indeed the case in general.
 The RHS of (\ref{uncer2}) can be zero only if $\vect{f}^\text{T}\vect{P}\vect{Q}\vect{f}=0$ where $\vect{P}=\vect{R}^\text{T}\vect{F}+\vect{F}^\text{T}\vect{R}$ and
 $\vect{Q}=\vect{F}^\text{T}\vect{F}-\left\langle{\vect{F}^\text{T}\vect{F}}\right\rangle$. This result is obtained upon localization of the integral in the definition
 of $\left\langle\vect{PQ}\right\rangle$, given $\Omega$ is arbitrary. Since $\vect{P}$ is positive definite and $\vect{Q}$ generally nonsingular, $\vect{PQ}$ is
 nonsingular too and thus $\vect{P}\vect{Q}\vect{f}\ne\vect{0}$. Clearly $\vect{PQ}$ is not skew symmetric and we may discount, almost surely, the other possibilities
 of $\vect{PQf}\perp\vect{f}$. This ensures, with probability 1, that the uncertainty relation is nontrivial. In other words, while eliminating the micro-rotation DOFs,
 they should be treated as stochastic processes.

Since, at a time instant, the mean of a micro-rotation DOF would typically be an order smaller than its translational counterpart, we may approximately identify
the micro-rotation DOF as a zero mean random variable. Retaining such noise terms in the final GLE is then crucial, as nonlocality necessarily implies nondeterminism \cite{propescu}, an aspect generally overlooked in nonlocal continuum theories.
The randomness in the micro-rotations, consequent upon the uncertainty relation Eqn.(\ref{uncer}), may be contrasted with that in the initial conditions due to thermal
fluctuations, yielding an additive noise term in the GLE. This leads us to recast Eqn.(\ref{10}) and write the GLE including the two noise sources as:
\begin{equation}\label{11}
\begin{split}
&\frac{d}{dt}\left\{ \begin{matrix}
   {{u}_{11}}  \\
   {{p}_{11}}  \\
\end{matrix} \right\}=\left[ \begin{matrix}
   0 & {1}/{{{m}_{11}}}\;  \\
   \left( {{{\hat{w}}}_{1111}}+{{{\overset{\lower0.5em\hbox{$\smash{\scriptscriptstyle\smile}$}}{w}}}_{0}}\right)&0\\\end{matrix}\right]\left\{\begin{matrix}{{u}_{11}}\\{{p}_{11}}\\\end{matrix}\right\}
+\\&\int\limits_0^t{\left\{\begin{matrix}0\\\hat{\eta}\left(s-t\right)\dot{u}_{11}\left(s\right)\end{matrix}\right\}ds}+\left\{\begin{matrix}0\\\xi\left(t\right)\end{matrix}\right\}
+\int\limits_0^t{\left\{\begin{matrix}0\\W_s^t{u}_{11}\left(s\right)\end{matrix}\right\}ds}
   \end{split}
\end{equation}
where $\hat{\eta}\left(\cdot\right)$ is the memory kernel that may be found from the second term on the right hand side of Eqn.(\ref{10}).
$\xi\left(t\right)$ is a linear combination of $u_{11}\left(0\right)$ and elements of $Y_b\left(0\right)$ at time $t$. An application of Lyapunov's central limit theorem (CLT) yields
$\xi\left(t\right)$ to be a zero mean Gaussian random variable at time $t$. On similar lines, the internal (multiplicative) noise $W_s^t$, arising from a weighted sum of the
micro-rotations as seen from the last term of Eqn.(\ref{10}),  may also be characterized as zero mean Gaussian for fixed $s$ and $t$. We may rewrite
 the new GLE (including an external forcing term $F\left(t\right)$ for completion) in its more familiar second order form.

 \begin{equation}\label{12}
 \begin{split}
 &m\ddot{u}+ku+\int_0^t{\eta\left(s-t\right)\dot{u}\left(s\right)ds}=\\&F\left(t\right)+\xi\left(t\right)+\int_0^t{W_s^t\left(s-t\right)u\left(s\right)ds}
 \end{split}
 \end{equation}

In Eqn.(\ref{12}), $m=m_{11}$ denotes the mass, $k=-\left(w_{1111}+{{{\overset{\lower0.5em\hbox{$\smash{\scriptscriptstyle\smile}$}}{w}}}_{0}}\right)$ the stiffness and $\eta=-\hat{\eta}$
the damping memory kernel for the system variable.
 Note that, if micropolar effects are not considered, i.e. $W_s^t$ and ${{{\overset{\lower0.5em\hbox{$\smash{\scriptscriptstyle\smile}$}}{w}}}_{0}}$ are identically zero, the usual form of GLE is retrieved.
\par \emph{A fluctuation-dissipation (FD) constraint.---}
In addition to the uncertainty constraint as in Eqn.(\ref{uncer}), the causality condition may impose a constraint on the new GLE. The latter would be similar to the FD theorem typically associated with the conventional GLE.
We designate the integral term on the right hand side of Eqn.(\ref{12}) by $y\left(t\right)$. A general scheme to represent $W_s^t$ could be via a Wiener chaos representation \cite{Wiener_chaos}, \cite{malliavin}.
However, for illustrative purposes, we consider the special form $W_s^t =\exp\left(\alpha\left(s-t\right)\right)\zeta\left(s\right)$ ($\alpha$ is some constant and
$\zeta\left(s\right)\sim\mathcal{N}\left(0,\sigma\left(s\right)\right)$) that enables representing $y\left(t\right)$ as the Markovian solution of a stochastic differential equation.
$\sigma$ is the intensity of $\zeta$. Thus Eqn.(\ref{12}) is equivalent to the coupled set of equations:

 \begin{equation}\label{13}
 m\ddot{u}+ku+\int_0^t{\eta\left(s-t\right)\dot{u}\left(s\right)ds}=F\left(t\right)+\xi\left(t\right)+y\left(t\right)
 \end{equation}
 \begin{equation}\label{14}
  \dot{y}\left(t\right)=y\left(t\right)+u\left(t\right)\zeta\left(t\right)
 \end{equation}

Under a random change of time $t\rightarrow\beta(t)$ \cite{r2}, $u\left(t\right)\zeta\left(t\right)$ may represented via a zero mean Brownian motion $\hat{B}\left(\beta\left(t\right)\right)$, where
 $\beta\left(t\right)=\int_0^t{\left(\sigma\left(s\right)^2u\left(s\right)^2\right)ds}$. Following the derivation of Kubo's second FDT \cite{kubo}, the constraint equation may now be directly written as

 \begin{equation}\label{15}
 \left\langle{\left(z\left(0\right)+\xi\left(0\right)\right)\left(z\left(t\right)+\xi\left(t\right)\right)}\right\rangle
 =\eta\left(t\right)k_BT
 \end{equation}
 where $z\left(t\right)=\left(\int_0^t{exp\left(s-t\right)\hat{B}\left(\beta\left(s\right)\right)ds}\right)$.

 Some manipulations on Eqn.(\ref{15}) would lead to Eqn.(\ref{16}).
   \begin{equation}\label{16}
\int_0^t{exp\left(s-t\right) \left\langle{\xi\left(0\right)\hat{B}\left(\beta\left(s\right)\right)}\right\rangle ds}=\eta\left(t\right)k_BT-\left\langle{\xi\left(0\right)\xi\left(t\right)}\right\rangle
 \end{equation}
 We see from Eqn.(\ref{16}) that, as the left hand side vanishes in absence of micropolarity, we get back the usual FD theorem.

\par
\begin{figure}[!htp]
    \begin{center}
            \label{1}
            \includegraphics[width=0.4\textwidth]{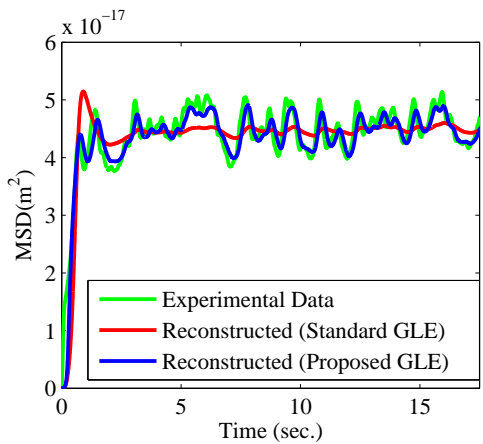}
    \end{center}
    \caption{%
Reconstruction of the MSDs via stochastic filter
     }%
     \label{inverse_both_GLE}
\end{figure}

 \emph{Numerical simulations and experiments.---}
The numerical tests are carried out by simulating the standard GLE and the proposed one within a Monte Carlo setup. In both the simulations Prony series approximations are adopted for characterizing the memory kernels and the non-Markovian SDEs are converted to a set of Markovian SDEs \cite{GLE_prony}. From the nonlocal interaction effects, as anticipated, the fluctuations in the steady-state regime are indeed captured in the simulation via the proposed GLE (Fig.\ref{simulation_GLE}), which parallels the
experimental observation (Fig.\ref{exp_GLE}). In Fig.\ref{simulation_GLE}, we also see that simulation using the
standard GLE fails to produce the steady-state fluctuations. The proposed GLE is further tested in the context of an inverse problem, wherein using it as a process model,
a stochastic projection on the experimental
MSD data through a nonlinear filter \cite{sarkar_physicad} leads to an estimate quite close to the measurement (Fig.\ref{inverse_both_GLE}). However, the same exercise with the standard GLE as process model
(for the same Monte Carlo sample size) produces a completely different response estimate.

\par On a concluding note we emphasize that a suitably designed internal noise accounts for nonlocality, a feature of space-separated multi-particle interaction,
within a 1-dimensional evolution equation. The new GLE can more faithfully model the response of the material taking into account some non-trivial aspects triggered by the microstructure.
In future work, other than working out a more general model for the internal noise, it would be of interest to see if similar
GLEs are derivable either from lattice models of materials or from Eshelby-type configurational forces.

\end{document}


\title{Supplementary Material for\\ ``Internal noise driven generalized Langevin equation from a nonlocal continuum model''}

\author{Saikat~Sarkar} \affiliation{Computational Mechanics Lab., Indian Institute of Science, Bangalore 560012, India}
\author{Shubhankar~Roy~Chowdhury} \affiliation{Computational Mechanics Lab., Indian Institute of Science, Bangalore 560012, India}
\author{Debasish~Roy} \affiliation{Computational Mechanics Lab., Indian Institute of Science, Bangalore 560012, India}
\author{Ram Mohan~Vasu} \affiliation{Instrumentation and Applied Physics, Indian Institute of Science, Bangalore 560012, India}

\noaffiliation
\vskip 0.25cm

\date{\today}
\maketitle

\section{Micropolar theory}
Micropolar theory \cite{eringen1}, a generalized continuum approach, provides a phenomenological route to include length scale effects
(generally manifested because of the micro-structural features of the material) in the response of a continuum
to external stimuli. As opposed to the classical continuum model where the motion of each material point is described by
its position vector alone, each such point in a micropolar continuum is additionally associated with a micro-structure which
can rotate rigidly independent of its surrounding material. Therefore, every material point within the micropolar framework is endowed with
six degrees of freedom (DOFs), namely three translational DOFs associated with the macro-element and three rotational DOFs with the micro-element. This
is contrasted with the classical description wherein only the former three are present. Along with this difference in the kinematic description,
the micropolar theory includes certain different features in the kinetic aspects too. The micropolar theory assumes the existence of a couple traction vector, in addition to the force traction considered in the classical continuum, for the force transfer mechanism across the interfaces of adjoining parts of the body.
That is why, unlike the classical symmetric Cauchy stress tensor, a micropolar continuum generates an asymmetric force stress and a couple
stress tensor. While the symmetric part of the force stress is responsible for the macro-deformation, the skew-symmetric portion along with the couple stress contribute to the rigid rotation of the micro-element. In the following three subsections, we briefly present the kinematics, equations of motion and constitutive relations for a micropolar continuum model.

\subsection{Kinematics}
Let ${\mathcal{B}_{0}}\subset {{\mathbb{R}}^{3}}$
be the reference configuration of a micropolar continuum body at time ${{t}_{0}}$. Macroscopically,
each material point $\vect{X}$ of this continuum deforms to $\vect{x}$ in its spatial configuration, $\mathcal{B}$, at time $t$. This translation is
represented by a smooth map $\chi $ as $\vect{x}=\chi \left( \vect{X},t \right)$. $\chi $ is a one-one and onto map; therefore possesses a unique inverse.
As already mentioned, the microstructure associated with each material point of such a continuum can rotate independently along with the macroscopic
deformation. To describe such rotational motion, assume an arbitrary vector $\boldsymbol{\Xi}$, at $\vect{X}$ in the reference configuration, which rotates to
$\vect{\xi}$ at $\vect{x}$ in its spatial configuration. In terms of a smooth invertible map $\zeta$, this motion is written as $\vect{\xi} =\zeta \left( \vect{\Xi} ,t \right)$.
The micro-rotation is described through a pseudo vector $\vect{\theta} ={{\left\{ \begin{matrix}
   {{\theta }_{1}} & {{\theta }_{2}} & {{\theta }_{3}}  \\
\end{matrix} \right\}}^{\text{T}}}$ with magnitude $\left| \vect{\theta}  \right|:=\varphi$. Associated with the pseudo vector $\vect{\theta}$, there exists a proper orthogonal tensor $\mathbf{R}$,
using which the rotation map of the microstructure is written as $\vect{\xi} =\mathbf{R}\left( \vect{\theta} ,t \right)\vect{\Xi}$. A closed form expression for $\mathbf{R}$
in terms of $\vect{\theta}$ and $\varphi$ is given by the Euler- Rodrigues formula:
\begin{equation}\label{s1}\mathbf{R}\left( \vect{\theta} ,t \right)=\cos \varphi \mathbf{I}+\frac{\sin \varphi }{\varphi }\mathbf{W}+\frac{1-\cos \varphi }{{{\varphi }^{2}}}\vect{\theta} \otimes \vect{\theta}\end{equation}
where $\mathbf{I}$ is the second order identity tensor, $\mathbf{W}$ the second order skew-symmetric tensor whose axial vector is $\vect{\theta} $ ($\mathbf{W}=-\varepsilon :\vect{\theta}$, $\varepsilon$ being
the third order alternate tensor or the Levi-Civita symbol) and $\otimes $ is the symbol for the exterior product or the dyadic product of vectors.
\subsubsection{Strain and Wryness Tensors}
Using the macro-translational motion $\chi $, the deformation gradient is defined as $\mathbf{F}=\frac{\partial \chi }{\partial \vect{X}}$. Similar
to the well known polar decomposition in the classical theory, $\mathbf{F}$ could be uniquely decomposed as $\mathbf{F}=\mathbf{RU}=\mathbf{VR}$. However a few
differences may be noted, viz. $\mathbf{R}$ here is not the macro-continuum rotation rather a measure of the micro-rotation, $\mathbf{U}$ and $\mathbf{V}$,
referred to as the right and left micropolar stretches respectively, are not symmetric unlike their classical counterparts. The right stretch tensor provides a material description
of stretch whereas the left stretch tensor gives the spatial description of the same. For the definitions of the micropolar strain tensor $\mathbf{E}$ and the wryness tensor $\mathbf{\Gamma }$ in a Lagrangian description, the following may be adopted:
$\mathbf{E}=\mathbf{U}-\mathbf{I}$ and $\mathbf{\Gamma }=-\frac{1}{2}\vect{\varepsilon} :\left( {{\mathbf{R}}^{\text{T}}}{{\nabla}_{\vect{X}}}\mathbf{R} \right)$.
The wryness tensor in the spatial configuration ($\boldsymbol{\gamma }$) is related to the material description as $\boldsymbol{\gamma }=\mathbf{R\Gamma }{{\mathbf{R}}^{\text{T}}}$.
\subsection{Equations of Motion}
The physical requirement of balances of linear and angular momenta for the continuum must be met. Localization of these global balance requirements leads to the following
equations of motion in the material description.
\begin{equation}\label{s2}{{\rho }_{0}}{{\left( \frac{{{\partial }^{2}}\chi }{\partial {{t}^{2}}} \right)}_{\vect{X}}}={{\nabla }_{\vect{X}}}\cdot \mathbf{T}+{{\rho }_{0}}{{b}_{0}} \;\;\;\;\;\text{and}\end{equation}
\begin{equation}\label{s3}\mathbf{J}{{\left( \frac{{{\partial }^{2}}\vect{\theta} }{\partial {{t}^{2}}} \right)}_{\vect{X}}}={{\nabla }_{\vect{X}}}\cdot \mathbf{M}+
{{\rho }_{0}}{{l}_{0}}-\vect{\varepsilon} :\mathbf{T}{{\mathbf{F}}^{\text{T}}}\end{equation}
Here $\mathbf{T}$ and $\mathbf{M}$ respectively refer to the force stress and couple stress tensors. ${{b}_{0}}$ and ${{l}_{0}}$ are the externally applied
body force and body couple densities respectively. ${{\rho }_{0}}$ and $\mathbf{J}$ respectively refer to the mass density and the micro-inertia tensor,
 all expressed in the material coordinate. The tensorial form of the micro-inertia $\mathbf{J}$ may be simplified to $\rho_0J\mathbf{I}$ in case of isotropic spin.
Solution of the above equations of motion along with proper boundary and initial conditions would represent the response of the continuum. Owing to $\vect{\theta}$ evolving on the special orthogonal nonlinear manifold SO(3), integration of Eqn.(\ref{s3}) is nontrivial. Specifically, the rotations of a material point at two different instances must be appropriately transformed to refer to the same tangent space of the manifold. However, for small $\vect{\theta}$, the necessary transformation is nearly the identity operator enabling integration within a purely linear space framework.
\subsection{Constitutive relations}
In the context of micropolar hyperelasticity, it is postulated that there exists a free energy functional (equivalently an internal strain energy functional
considering only the mechanical response instead of the thermo-mechanical one) which defines the constitutive relations given as
$\mathbf{T}={{\rho }_{0}}\frac{\partial \psi \left( \mathbf{U},\mathbf{\Gamma } \right)}{\partial {{\mathbf{U}}^{\text{T}}}}$ and $\mathbf{M}={{\rho }_{0}}\frac{\partial \psi \left( \mathbf{U},\mathbf{\Gamma } \right)}
{\partial {{\mathbf{\Gamma }}^{\text{T}}}}$.
Material description of the free energy functional is used to establish these relations. The specific form of the free energy density for the non-linear continuum considered here is given below.
\begin{equation}\label{s4}\psi \left( \mathbf{U},\mathbf{\Gamma },\mathbf{R} \right)={\bar{\psi }}\left( \mathbf{U},\mathbf{R} \right)+\hat{\psi }\left( \mathbf{U},\mathbf{R} \right)+\tilde{\psi }\left( \mathbf{\Gamma },\mathbf{R} \right)\end{equation}

where $${{\rho }_{0}}{\bar{\psi }}\left( \mathbf{U},\mathbf{R} \right)=\frac{1}{2}\tilde{\mu }\left( tr\left( \mathbf{RU}{{\mathbf{U}}^{\mathbf{T}}}{{\mathbf{R}}^{\text{T}}} \right)-3 \right)+
h\left( \det \left( \mathbf{F} \right) \right)$$
\[{{\rho }_{0}}\hat{\psi }\left( \mathbf{U},\mathbf{R} \right)=\frac{1}{4}\tilde{\kappa }\left( tr\left( \mathbf{RU}{{\mathbf{U}}^{\mathbf{T}}}{{\mathbf{R}}^{\text{T}}} \right)
-tr\left( \mathbf{R}{{\mathbf{U}}^{2}}{{\mathbf{R}}^{\text{T}}} \right) \right)\]
\[{{\rho }_{0}}\tilde{\psi }\left( \mathbf{\Gamma },\mathbf{R} \right)=\frac{1}{2}\left( \tilde{\alpha }{{\left( tr\left( \mathbf{R\Gamma }{{\mathbf{R}}^{\text{T}}} \right) \right)}^{2}}
+\tilde{\beta }tr\left( \mathbf{R}{{\mathbf{\Gamma }}^{2}}{{\mathbf{R}}^{\text{T}}} \right)+\tilde{\gamma }tr\left( \mathbf{R\Gamma }{{\mathbf{\Gamma }}^{\text{T}}}{{\mathbf{R}}^{\text{T}}} \right) \right)\]
The function $h$ above represents an energy density associated with volumetric deformation, the functional form of which is
$h=\frac{{\tilde{\lambda }}}{4}\left( \det {{\left( \mathbf{F} \right)}^{2}}-1 \right)-\frac{{\tilde{\lambda }}}{2}\ln \left( \det \left( \mathbf{F} \right) \right)-\tilde{\mu }\ln
\left( \det \left( \mathbf{F} \right) \right)$.
Here $\tilde{\mu }$, $\tilde{\lambda }$, $\tilde{\kappa }$, $\tilde{\alpha }$,  $\tilde{\beta }$ and $\tilde{\gamma }$ are the material parameters to be determined from experiments.
\section{Formulation of discrete Hamiltonian }
The Hamiltonian ($H$) in a closed system may be given in terms of the total energy ($\Pi$) comprised of the kinetic energy and the strain energy stored in the body.
The kinetic energy of the body is given by $K.E.=\frac{1}{2}\int_{{{V}_{0}}}{{{\rho }_{0}}\left( \vect{v}.\vect{v}+J\vect{\omega} .\vect{\omega} \right)d\vect{V}}$ and the potential energy stored is given by
$P.E.=\int_{{{V}_{0}}}{{{\rho }_{0}}\psi d\vect{V}}$. These volume integrals are defined on the reference configuration where $d\vect{V}$ denotes a volume element
and ${{V}_{0}}$ the volume of the configuration ${{B}_{0}}$. $\vect{v}={{\left( \frac{\partial \chi }{\partial t} \right)}_{\vect{X}}}$ is the linear velocity and
$\vect{\omega}={{\left( \frac{\partial \vect\theta }{\partial t} \right)}_{\vect{X}}}$ the spin velocity in the reference coordinate. Thus
$H=\frac{1}{2}\int_{{{V}_{0}}}{{{\rho }_{0}}\left( \vect{v}.\vect{v}+J\vect{\omega} .\vect{\omega} \right)d\vect{V}}+\int_{{{V}_{0}}}{{{\rho }_{0}}\psi d\vect{V}}$.
As we consider micro-rotations and their gradients to be small, we may ignore the contribution of $\tilde{\psi }$ in $H$. Moreover, the material considered (e.g. polymer)
here is assumed to be elastic with nearly isochoric ($\det \left( \mathbf{F} \right)\approx 1$) deformation. Hence, we may ignore the contribution of $h$ too. Thus we arrive at a simpler description of H given as:
\begin{equation}\label{s5}H=\frac{1}{2}\int_{{{V}_{0}}}{{{\rho }_{0}}\left( \vect{v}.\vect{v}+J\vect{\omega} .\vect{\omega} \right)d\vect{V}}+\int_{{{V}_{0}}}{\left( \frac{1}{2}\tilde{\mu }\left( tr\left( \mathbf{F}{{\mathbf{F}}^{\text{T}}} \right)-3
\right)+\frac{1}{4}\tilde{\kappa }\left( tr\left( \mathbf{F}{{\mathbf{F}}^{\text{T}}} \right)-tr\left( \mathbf{F}{{\mathbf{R}}^{\mathbf{T}}}\mathbf{F}{{\mathbf{R}}^{\mathbf{T}}} \right) \right) \right)d\vect{V}}\end{equation}
Towards writing $H$ in terms of the displacement vector, $\mathbf{u}=\chi \left( \vect{X} \right)-\vect{X}$, we use the relation, $\mathbf{F}={{\nabla }_{\vect{X}}}\mathbf{u}+\mathbf{I}$ and rewrite Eqn.(\ref{s5}) as
(for conciseness $\nabla $ is used instead of ${{\nabla }_{\vect{X}}}$).

\begin{equation}\label{s6}H=\frac{1}{2}\int_{V_0}{\rho_0\left(\vect{v}.\vect{v}+J\vect{\omega}.\vect{\omega}\right)}+
\int_{V_0}\left(\begin{split}&\frac{1}{2}{\tilde{\mu}\,tr\left(\nabla\mathbf{u}\left(\nabla\mathbf{u}\right)^\mathbf{T}+\nabla\mathbf{u}+\left(\nabla\mathbf{u}\right)^\mathbf{T}\right)}+\\
&{\frac{1}{4}\tilde{\kappa}\,tr\left(\nabla\mathbf{u}\left(\nabla\mathbf{u}\right)^\mathbf{T}+\nabla\mathbf{u}+\left(\nabla\mathbf{u}\right)^\mathbf{T}+3\right)}-\\
&\frac{1}{4}\tilde{\kappa}\,tr\left(\nabla\mathbf{uR^T}\nabla\mathbf{uR^T}+\nabla\mathbf{uR^{T}R^{T}}+\mathbf{R^T}\nabla\mathbf{u}\mathbf{R^T}+\mathbf{R^T}\mathbf{R^T}\right)\end{split}\right)d\vect{V}\end{equation}
In what follows, we describe a consistent discretization of $H$ given in Eqn.(\ref{s6}). Using the particle mass $m_{i\alpha}={\rho_0}_\alpha\Delta\vect{V}_\alpha$ and the micro-inertia $I_{i\alpha}=\left(\rho_0 J\right)_\alpha\Delta\vect{V}_\alpha$ for $\alpha=1,...,N$, the following discrete form of the kinetic energy
may be arrived at.
\begin{equation}\label{k1}
\frac{1}{2}\sum\limits_{\alpha=1}^{N}{\left(m_{i\alpha}v_{i\alpha}^2+I_{i\alpha}\omega_{i\alpha}^2\right)}
=\frac{1}{2}\sum\limits_{\alpha=1}^{N}{\left(\frac{p_{i\alpha}^2}{m_{i\alpha}}+\frac{p_{\theta i\alpha}^2}{I_{i\alpha}}\right)}
\end{equation}
 $N$ is the total number of material particles. In $m_{i\alpha}$ and $I_{i\alpha}$, the subscript $i$ is introduced to maintain notational similarity with the other vector field variables (e.g. $v_{i\alpha}$) and this is helpful in using Einstein's
summation convention in $i$.
In order to discretize the potential energy, a kernel approximation of the the displacement field is adopted. For example, we may write
$$\int_{{{V}_{0}}}{\left( \,tr\left( \nabla \mathbf{u}{{\left( \nabla \mathbf{u} \right)}^{\text{\bf{T}}}} \right) \right)d\vect{V}}\approx
\int_{{{V}_{0}}}{\left\{ \int_{{{V}_{0}}}{{{\left[ {{\nabla }_{{\vect{X}'}}}\Phi \left( \vect{X}-\vect{X}' \right) \right]}_{i}}{{{u_{j}'}}}d\vect{V}'} \right\}\left\{ \int_{{{V}_{0}}}
{{{\left[ {{\nabla }_{{\vect{X}'}}}
\Phi \left( \vect{X}-\vect{X}' \right) \right]}_{j}}{{{u_{i}'}}}d\vect{V}'} \right\}d\vect{V}},$$
where $\Phi \left( y \right)$ is a sufficiently smooth symmetric kernel with respect to $y$.
${{u_{j}'}}$ is the $j^{\text{th}}$ component of $\mathbf{u}\left( {\vect{X}'} \right)$. A subsequent particle representation leads to the following expression:
\begin{equation}\label{s7}
\begin{split}
   \int_{{{V}_{0}}}{\left( \,tr\left( \nabla \bf{u}{{\left( \nabla \bf{u} \right)}^{\text{\bf{T}}}} \right) \right)d\vect{V}}
    \approx\sum\limits_{\gamma =1}^{N}{\left( \sum\limits_{\alpha =1}^{N}{{{\left\{ \left( {{\nabla }_{{\vect{X}'}}}\Phi  \right)_{i}^{\gamma }{{u}_{j}} \right\}}_{\alpha }}\Delta {\vect{V}_{\alpha }}}
 \sum\limits_{\beta =1}^{N}{{{\left\{ \left( {{\nabla }_{{\vect{X}'}}}\Phi  \right)_{j}^{\gamma }{{u}_{i}} \right\}}_{\beta }}\Delta {\vect{V}_{\beta }}} \right)}\Delta {{V}_{\gamma }}
 ={\sum\limits_{\alpha,\beta =1}^{N}{{{w}_{ij\alpha \beta }}{{u}_{j\alpha }}{{u}_{i\beta }}}} \\
\end{split}
\end{equation}
Here, ${{u}_{m\tau }}$ denotes the $m^{\text{th}}$ component of $\bf{u}$ evaluated at the $\tau^{\text{th}}$ material particle.
\\${{w}_{ij\alpha \beta }}=\sum\limits_{\gamma =1}^{N}{\left( {{\left\{ \left( {{\nabla }_{{{X}'}}}\Phi  \right)_{i}^{\gamma } \right\}}_{\alpha }}
{{\left\{ \left( {{\nabla }_{{{X}'}}}\Phi  \right)_{j}^{\gamma } \right\}}_{\beta }}\Delta {{V}_{\alpha }}\Delta {{V}_{\beta }} \right)}\Delta {{V}_{\gamma }}$ is the appropriate weight.
Now, for small $\vect{\theta}$, we use the approximation $\mathbf{R}\approx\mathbf{I}-\boldsymbol{\varepsilon}:\vect{\theta}$ to get:
\begin{equation}\label{s8}
\begin{split}
   &\int_{{{V}_{0}}}{ \,tr\left( \nabla \bf{u}{{\left(\bf{RR} \right)}^{\text{\bf{T}}}} \right) d\vect{V}}
   =\int_{{{V}_{0}}}{ \,tr\left( \mathbf{R^T}\nabla \bf{u}{{\bf{R} }^{\text{\bf{T}}}} \right)d\vect{V}}
   \approx{\sum\limits_{\alpha,\gamma =1}^{N}{{\tilde{w}_{i\alpha \gamma }}{u_{j\alpha }}\left(\left(\theta_{j\gamma}\theta_{i\gamma}+2\text{S}\left(\theta_{k\gamma}\right)_{k\ne j,i}\right)_{i\ne j}+
   \left(1-\sum\limits_{k\ne i}{\theta_{k\gamma}^2}\right)_{i=j}\right)}} \\
\end{split}
\end{equation}
where ${\tilde{w}_{i\alpha \gamma }}=\sum\limits_{\gamma =1}^{N}{\left( {{\left\{ \left( {{\nabla }_{{{X}'}}}\Phi  \right)_{i}^{\gamma } \right\}}_{\alpha }}
\Delta {{V}_{\alpha }} \right)}\Delta {{V}_{\gamma }}$. For $\left(j,i\right)=\left(1,2\right), \left(2,3\right), \left(3,1\right)$, $\text{S}=1$ and for other combinations of $\left(j,i\right)$, $\text{S}=-1$.
Neglecting contributions from $\,tr\left(\nabla\mathbf{uR^T}\nabla\mathbf{uR^T}\right)$ and $\,tr\mathbf{\left(RR\right)^T}$, the discrete Hamiltonian finally takes the following form.
\begin{equation}\label{s9}
\begin{split}
 H=&\frac{1}{2}\sum\limits_{\alpha=1}^{N}{\left(\frac{p_{i\alpha}^2}{m_{i\alpha}}+\frac{p_{\theta i\alpha}^2}{I_{i\alpha}}\right)}+
 {\sum\limits_{\alpha,\beta =1}^{N}{{{w}_{ij\alpha \beta }}{{u}_{j\alpha }}{{u}_{i\beta }}}}+
 {\sum\limits_{\alpha,\gamma =1}^{N}{{\tilde{w}_{i\alpha \gamma }}{u_{j\alpha }}\left(\left(\theta_{j\gamma}\theta_{i\gamma}+2\text{S}\left(\theta_{k\gamma}\right)_{k\ne j,i}\right)_{i\ne j}
   +\left(1-\sum\limits_{k\ne i}{\theta_{k\gamma}^2}\right)_{i=j}\right)}}
\end{split}
\end{equation}

\section{Equations of motion of the discrete system}
The governing dynamics for the system and bath variables, described through Hamilton's equations in terms of the displacement DOFs and the momenta, are given by the equation
pairs (\ref{c1}, \ref{2}) and (\ref{3}, \ref{4}) respectively.
\begin{equation} \label{c1}
\frac{d}{dt}\left( {{u}_{11}} \right)=\frac{\partial H}{\partial {{p}_{11}}}=\frac{{{p}_{11}}}{{{m}_{11}}}
\end{equation}

\begin{equation}\label{2}\begin{split}{&\frac{d}{dt}\left(p_{11}\right)=-\frac{\partial H}{\partial u_{11}}={\hat{w}}_{1111}{u}_{11}+\sum\limits_{\beta=2}^N{w_{i11\beta}u_{i\beta}}+
\sum\limits_{\gamma=1}^N{\overset{\lower0.5em\hbox{${\scriptscriptstyle\frown}$}}{w}_{i11\beta}\left[\left(\theta_{1\gamma}\theta_{i\gamma}+
2\text{S}\left(\theta_{k\gamma}\right)_{k\ne{j},i}\right)_{i\ne1}
+\left(\theta_{k1}^2\right)_{i=1}\right]}}\end{split}\end{equation}
\begin{equation}\label{3}\frac{d}{dt}\left( {{{\hat{u}}}_{\theta }} \right)=\frac{\partial H}{\partial {{{\hat{p}}}_{u\theta }}}={{\left\{ \begin{matrix}
   \frac{{{p}_{21}}}{{{m}_{21}}} & ... & \frac{{{p}_{3N}}}{{{m}_{3N}}} & \frac{{{p}_{\theta 11}}}{{{I}_{11}}} & ... & \frac{{{p}_{\theta 3N}}}{{{I}_{3N}}}  \\
\end{matrix} \right\}}^{\text{T}}}\end{equation}
\begin{equation}
\label{4}\frac{d}{dt}{\hat{p}_{u\theta}}=-\frac{\partial H}{\partial \hat{u}_\theta}
                          =\left[\begin{split}\sum\limits_{\beta=1}^N{w_{i12\beta}u_{i\beta}}&+\sum\limits_{\gamma=1}^N{\overset{\lower0.5em\hbox{${\scriptscriptstyle\frown}$}}{w}_{i2\gamma}L_{\theta 12\gamma}}\\
  &\vdots\\
   \sum\limits_{\beta=1}^N{w_{i3N\beta}u_{i\beta}}&+\sum\limits_{\gamma=1}^N{\overset{\lower0.5em\hbox{${\scriptscriptstyle\frown}$}}{w}_{iN\gamma}L_{\theta 3N\gamma}}\\
   \mathbf{C_{\boldsymbol{\theta}}}\left(u_{11} ... u_{3N}\right)^{\text{T}}&+\mathbf{D}\left(u_{11} ... u_{3N}\right)^{\text{T}}\end{split}\right]
  \end{equation}
where $L_{\theta 12\gamma}=\left[\left(\theta_{1\gamma}\theta_{i\gamma}+2\text{S}\left(\theta_{k\gamma}\right)_{k\ne{j},i}\right)_{i\ne1}+\left(\theta_{k2}^2\right)_{i=1}\right]$ and
$L_{\theta 3N\gamma}=\left[\left(\theta_{3\gamma}\theta_{i\gamma}+2\text{S}\left(\theta_{k\gamma}\right)_{k\ne{j},i}\right)_{i\ne1}+\left(\theta_{kN}^2\right)_{i=3}\right]$.
 ${{\hat{u}}_{\theta }}={{\left\{ \begin{matrix}
   {\hat{u}}^{\text{T}} & {\theta}^{\text{T}}   \\
\end{matrix} \right\}}^{\text{T}}}={{\left\{ \begin{matrix}
   {{u}_{21}} & ... & {{u}_{3N}} & {{\theta }_{11}} & ... & {{\theta }_{3N}}  \\
\end{matrix} \right\}}^{\text{T}}}$consists of displacements and micro-rotations of bath variables and  ${{\hat{p}}_{u\theta }}={{\left\{ \begin{matrix}
   {{p}_{21}} & ... & {{p}_{3N}} & {{p}_{\theta 11}} & ... & {{p}_{\theta 3N}}  \\
\end{matrix} \right\}}^{\text{T}}}$, the corresponding linear and angular momenta. $\mathbf{C}_{\boldsymbol{\theta}}$ is a matrix each of whose elements is a linear combination of $\theta_{i\alpha}$
where $i\in \left\{ 1,2,3 \right\}$ and $\alpha \in \left\{ 1,2,...,N \right\}$. $\mathbf{D}$ is a matrix of constant coefficients. $\hat{w}_{1111}$ and
$\overset{\lower0.5em\hbox{${\scriptscriptstyle\frown}$}}{w}_{i11\beta}$ are constants related respectively to  $w_{1111}$ and ${w}_{i11\beta}$ through the constant multipliers resulting from the partial derivatives of the Hamintonian. A similar definition holds for the constants $\overset{\lower0.5em\hbox{${\scriptscriptstyle\frown}$}}{w}_{i\alpha\gamma}$.